# The study of uniaxial-biaxial phase transition of confined hard ellipsoids using density functional theory

M. Moradi[*], B. Binaei Ghotbabadi[†]

*Department of Physics, College of Science, Shiraz University, Shiraz 71454, Iran*

R. Aliabadi[‡]

*Department of Physics, College of Science, Fasa University, Fasa, Iran*

[*]*moradi@susc.ac.ir*

[†]*binaei737@gmail.com*

[‡]*aliabadi@fasau.ac.ir*

The density profiles and corresponding order parameters of the hard ellipsoids confined between two hard walls and also in contact with a single hard wall are studied using the density functional theory. The Hyper-Netted Chain (HNC) approximation is used to write excess grand potential of the system with respect to the bulk value. To simplify the calculations we use restricted orientation model (ROM) for the orientation of ellipsoids to find the density profiles and order parameters. Density functional theory shows that there is a uniaxial-biaxial phase transition near a single hard wall and also between two hard walls for a fluid consisting of uniaxial hard ellipsoidal particles with finite elongation.

## 1. Introduction

To understand the physical properties of the different states of matter, various theoretical methods like integral equations[1] and molecular dynamics[2] have been used. One of the most successful theories is the density functional theory (DFT) which has been proven to be a powerful tool in the study of the thermodynamics and structures of the bulk and non-uniform phases of the molecular fluids [3-5]. It has been applied to examine, nuclei, atoms, molecules, solids and the quantum and classical fluids. For instance, this theory has been applied to investigate the thermodynamics of inhomogeneous hard monodisperse and polydisperse systems like elliptical and ellipsoidal liquid crystals [3,6], Gay-Berne film[7], disc-sphere mixture[8], hard circular cylinders and hard Gaussian overlap fluids[9, 10], dipolar hard spheres between two hard walls[11], the electronic structure of the matter[12] and the semiconductors[13]. Interestingly, DFT is also very successful in the study of social dynamics such as vaccination[14] and human cooperation[15].

Confinement makes the phase behavior of a liquid crystal (LC) much richer[16, 17]. The confined LCs have very important role in the LC displays, biosensors, the nanotechnology industry and etcetra[18]. This is why they have received remarkable attention during the years. Cheung and Schmidt[19] studied a system of soft ellipsoidal molecules confined between two planner walls using classical DFT where both the isotropic and nematic phases were considered. They evaluated the excess free energy using two different *ansatze* and the intermolecular interaction was incorporated using two different direct correlation function (DCF). The calculated densities and order parameters have been compared with simulation results of the same system. Varga et al.[20] used a density functional approach to describe the orientational ordering of nonpolar and dipolar Gay-Berne fluids. Moradi et al.[21] obtained the interaction forces between nano-circular particles suspended in a hard-ellipse fluid.

Han Miao and Hong-Ru Ma[22] studied 3D hard prolate ellipsoids with various densities and aspect ratios confined between two walls and orientational order parameters in the bulk fluid at the same density using Monte Carlo simulations. They concluded that the anisotropic ordering near the wall is induced by the confinement effect of the walls.

Because of strong competition between wall-molecule and molecule-molecule forces[23-25], anisotropic hard particles near a substrate show interesting phases such as uniaxial and biaxial (U-B) nematic, discotic and spatially ordered phases. The biaxial phases have been predicted and observed in computer simulation of hard ellipsoids. McBride and Lomba [26] investigated such a system and found that the elongated uniaxial models indicate the first-order nature of the isotropic-nematic transition for various ellipsoidal models. In their work, it was obtained the formation of a biaxial phase. This phase has opened a new area of research. In the biaxial phase, the system has three different optical axes that give rise to the diversity of physical properties, whereas there is only one preferred optical axis in the uniaxial nematic. For example, Aliabadi et al.[27] examined the orientation ordering of confined hard rectangular rods using the second virial density functional theory and the effects of varying shape anisotropy on the U-B, wetting, and isotropic-nematic transitions.

In this paper, we focus theoretically on the presence of U-B phase transition of a monodisperse hard ellipsoidal LC with finite aspect ratios near a single hard wall and confined between two parallel hard walls applying the DFT in the HNC approximation and the Zwanzig three-state model where the particles are allowed to orient along x, y and z-axes. In contrast to the simplicity of this model, it gives the basic physics of the systems and also makes the numerical calculations much simpler [28] and does not alter qualitatively the results for coexisting densities and order parameters, i.e., the results are in good agreements with the theoretical predictions earned from freely rotating hard rod particles [24,27,29]. This article is organized as follows: In Sec. 2 we describe DFT excess grand potential in the presence of hard planar wall(s). In Sec. 3 we show and discuss our numerical results. Finally, in Sec. 4 we summarize and conclude the remarkable results. The direct correlation function (DCF) of hard ellipsoidal particles is presented in the Appendix.

## 2. Grand potential

We study the ordering properties of hard ellipsoids using DFT. As shown in Fig. 1, these particles with major axis *2a* and minor axis *2b* are confined by hard walls. For a model fluid containing hard ellipsoidal molecules in the presence of an external potential $V(\vec{r},\omega)$, the excess grand potential with respect to its bulk is a unique functional of number density. Up to second order in density and using Hyper-Netted Chain (HNC) approximation the excess grand potential is given by [9, 11]

$$\beta\Delta\Omega[\rho] = \int d\vec{r}\, d\omega \rho(\vec{r},\omega)\left[\ln\frac{\omega_T \rho(\vec{r},\omega)}{\rho_b} - 1\right] + \int d\vec{r}\, d\omega \rho(\vec{r},\omega) V_{ext}(\vec{r},\omega)$$
$$-\frac{1}{2}\int d\vec{r}_1\, d\omega_1\, d\vec{r}_2\, d\omega_2 C(\vec{r}_{12},\omega_1,\omega_2)\left[\rho(\vec{r}_1,\omega_1) - \frac{\rho_b}{\omega_T}\right]\left[\rho(\vec{r}_2,\omega_2) - \frac{\rho_b}{\omega_T}\right]$$
(1)

where $C(\vec{r}_{12},\omega_1,\omega_2)$ is the direct correlation function (DCF) of a homogeneous molecular fluid and $\beta = 1/k_B T$, $\omega_i \equiv \theta_i, \varphi_i$ denotes molecular orientation and $\rho(\vec{r},\omega)$ is the number density of this fluid at the point $\vec{r}$ with the molecular orientation $\omega \equiv \theta, \varphi$  $\omega_T$ is the total solid angle available to the molecules and $\rho_b$ is the bulk density of the fluid that it can be written as follows in term of the chemical potential, $\mu$,

$$\rho_b = \exp(\beta\mu) \tag{2}$$

We consider a system of hard ellipsoids confined between parallel hard walls and choose the z-axis normal to the wall. Since the center of the molecules can move freely between the walls, according to the restricted orientation model (ROM), [9] these molecules are aligned only in six particular directions $\pm x$, $\pm y$, and $\pm z$. The excess grand potential per unit area with respect to its bulk value can be written as Eq.(3).

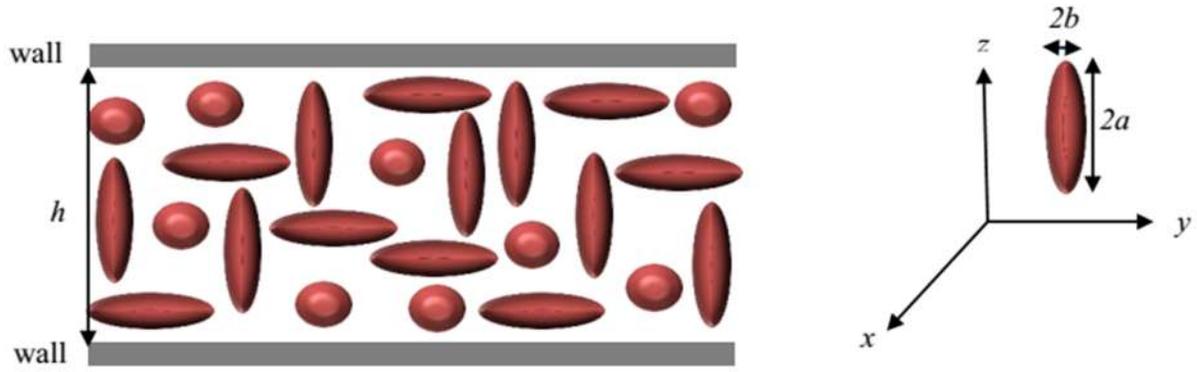

Fig. 1: Schematic representation of the hard ellipsoids confined between two hard walls with wall separation h. According to ROM model, molecules are allowed to be along the x, y and z-axes. *2a* and *2b* are the major and minor axes of a molecule, respectively.

$$\beta\Delta\Omega_A = \sum_\alpha \int dz\, \rho_\alpha(z)(\ln\frac{6\rho_\alpha(z)}{\rho_b} - 1) + \beta\sum_\alpha \int dz\, \rho_\alpha(z)\, V(z,\omega_\alpha) - \tag{3}$$

$$-\frac{1}{2}\sum_{\alpha,\beta}\int dz_1 dz_2\, C_{\alpha\beta}(z_1 - z_2) \times [\rho_\alpha(z_1) - \frac{\rho_b}{6}][\rho_\beta(z_2) - \frac{\rho_b}{6}]$$

where

$$C_{\alpha\beta}(z_1 - z_2) = \iint dx_2 dy_2\, C_{\alpha,\beta}(\vec{r}_{12}, \omega_{1\alpha}, \omega_{2\beta}) \tag{4}$$

This DCF (Eq. 4) depends on the orientation of the molecules; indices $\alpha$, and $\beta$ show this orientation.

We obtain the equilibrium density profiles from the functional minimization of Eq. (3) with respect to the density $\rho_\alpha(z)$. The coupled integral equations of directional density profiles can be obtained as:

$$\rho_\alpha(z_1) = \frac{\rho_b}{6}\exp\{-\beta V(z_1,\omega_\alpha) + \sum_\beta \int dz_2 C_{\alpha\beta}(z_1-z_2)\left[\rho_\beta(z_2)-\frac{\rho_b}{6}\right]\} \quad (\alpha,\beta = \pm x, \pm y, \pm z) \quad (5)$$

In these cases, the density profiles are only the functions of $z$ variable as well as the number densities have <u>nonzero</u> values between the walls and zero everywhere. By applying the existent symmetries, they can be written as

$$\rho_x(z) = \rho_{-x}(z), \qquad \rho_y(z) = \rho_{-y}(z), \qquad \rho_z(z) = \rho_{-z}(z) \quad (6)$$

where $\rho_{\pm x}(z), \rho_{\pm y}(z), \rho_{\pm z}(z)$ represent the density profiles of the molecules along $\pm x, \pm y, \pm z$ directions, respectively. Direct correlation functions are used to solve the coupled integral equations and find the density profiles. These DCFs are symmetric, i.e.,

$$C_{\pm\alpha,\pm\beta}(z) = C_{\alpha,\beta}(z) \quad (7)$$

$$C_{13}(z) = C_{31}(z) = C_{32}(z) = C_{23}(z), \qquad C_{11}(z) = C_{22}(z)$$

The notation $\pm 1, \pm 2, \pm 3$ are used here respectively for six particular directions $\pm x, \pm y,$ and $\pm z$.

By solving the above equations and applying the boundary conditions, we can obtain the density profiles of confined hard ellipsoids. These boundary conditions depend on the orientation of the molecules, so we can write the number density of the molecules parallel to the walls, $\rho_x, \rho_y$ and the molecules perpendicular to the walls, $\rho_z$, with wall separation $h$ as follows

$$\begin{cases} \rho_x = 0, \quad \rho_y = 0 & \text{for} \quad |z| > \left(\dfrac{h}{2} - b\right) \\ \rho_z = 0 & \text{for} \quad |z| > \left(\dfrac{h}{2} - a\right) \end{cases} \qquad (8)$$

Since these particles are confined between two parallel hard walls, the density profiles are zero outside the walls. As explained, the number density profiles can be obtained using boundary conditions, Eq. (8) and the DCF of the fluid which is introduced in the Appendix. We define three different orientational order parameters to determine the degree and the type of ordering. By choosing the $x$-axis for the director, the orientation average of the second Legendre polynomial $S_x$ can be obtained as

$$S_x = \frac{\rho_x - \rho_y/2 - \rho_z/2}{\rho} \qquad (9)$$

and two other order parameters $\Delta_{xy}$ and $\Delta_{yz}$, which evaluate the degree of biaxiality between $x$ and $y$ orientations and between $y$ and $z$ which are defined as[27]

$$\Delta_{xy} = \frac{\rho_x - \rho_y}{\rho} \qquad (10)$$

$$\Delta_{yz} = \frac{\rho_y - \rho_z}{\rho} \qquad (11)$$

where $\rho = \rho_x + \rho_y + \rho_z$ is the total density profile.

In all the figures, we have used dimensionless $\rho_i^* = \rho_i (2b)^3$, $Z^* = Z/2b$ and $h^* = h/2b$.

## 3. Results and Discussion

In this section, we present our numerical results for hard ellipsoids near a hard wall and confined in a slit pore. We solve three self-consistent and the coupled integral Eqs. (5) by the standard iteration method at given packing fraction ($\eta=4\pi ab^2\rho_b/3$) and the chemical potential (μ). The density profiles and the corresponding order parameters of ellipsoids with elongation $k=$ *2a/2b=3* in contact with a single hard wall are plotted in Fig. 2, for three different chemical potentials. At the wall, the density profiles and order parameters have different behaviors. For all the chemical potentials, we see a sharp and high first peak of $\rho^*_x$ at the wall and damped oscillatory behavior near the wall. Far from the wall, they approach to the bulk density.

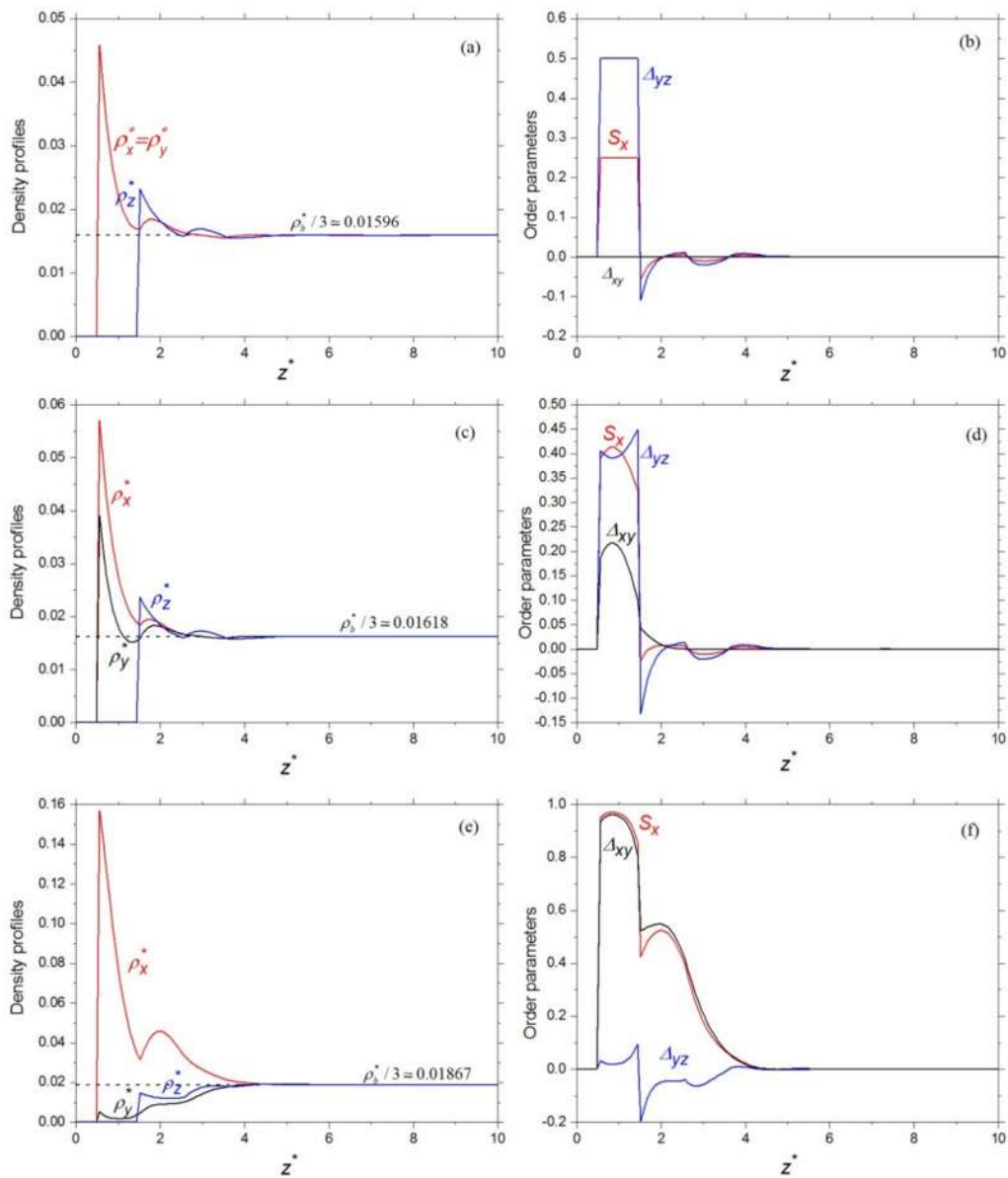

Fig. 2. Density profiles and order parameters of the hard ellipsoid particles in contact with a single hard wall. The elongation is $2a/2b=3$. Dotted lines indicate the bulk component densities. (a) and (b) are plotted for reduced chemical potential $\beta\mu=-2.4$, where the phase is isotropic even near the wall. (c) and (d) obtained at $\beta\mu=-2.33$, this phase is biaxial isotropic where there is a difference between the densities of x and y-axes near the wall but the phase is isotropic far from the wall. (e) and (f) show the results of $\beta\mu=-2.177$ where most of the particles have aligned with the x-axis as it can be seen from $S_x$ in Fig. (f).

In this area, the order parameters are close to zero for all the chemical potentials and the phase is isotropic. At $\beta\mu=-2.4$ the phase is isotropic i.e., $\rho_x^* = \rho_y^*$ ($\Delta_{xy}=0$) (Fig. 2. (a), (b)). By increasing the chemical potential ($\beta\mu=-2.33$), the biaxial order starts to appear ($\rho_x^* \neq \rho_y^*$) near the hard wall as it has been shown in Figs. 2(c) and 2(d). In this case the number of particles that tend to be parallel to the x-axis increases. This phase is biaxial ($\Delta_{xy}\neq 0$, $\Delta_{yz}\neq 0$). As it can be seen from Figs. 2(e) and 2(f), near the wall, an increment in the chemical potential ($\beta\mu=-2.177$) gives rise to higher $\rho_x^*$ and smaller $\rho_y^*$ and $\rho_z^*$ with respect to $\rho_b^*/3$, so the thickness of the nematic layer grows ($\rho_x^* \gg \rho_y^*, \rho_z^*$). This Fig. shows that at the larger chemical potential, the value of $\rho_y^*$ and $\rho_z^*$ almost are closed to each other so $\Delta_{yz}$ becomes smaller. In summary, Fig. (2) indicates the presence of the U-B transition induced by the wall as well as the formation of a nematic layer near the wall that its thickness increases with increasing the chemical potential. The large first peaks in the densities show the strong adsorption between the wall and the molecules. In addition to these, the waves in densities depict there is a layering fluid near the wall.

The density profiles and corresponding order parameters of the confined fluid between two parallel hard walls have been exhibited in Fig. 3 which obtained based on the numerical solution of Eq. (5) for $2a/2b=3$ and three different chemical potentials, at the lowest chemical potential ($\beta\mu=-3$) x, and y orientations are equivalent ($\rho_x^* = \rho_y^*$) and the phase is isotropic. In this case, $\Delta_{xy}=0$ while two other order parameters are zero in the middle of the pore (Fig. 3 (a), (b)). By increasing the chemical potential ($\beta\mu=-2.33$), the biaxial isotropic phase appears. As Fig. 3 (c) shows the density profiles are different in x, and y orientations near the walls except for the middle of the pore. It is clear from Fig. 3(d), except in the central region, each three order parameters are non-zero everywhere ($\Delta_{xy}\neq 0$, $\Delta_{yz}\neq 0$, $S_x\neq 0$). At the larger chemical potential ($\beta\mu$=-2.1), most of the particles parallel to the direction of the *x*-axis. As a result, they form a nematic film that exists even in the central region, this is due to both the walls strengthen the biaxial order. Here $\Delta_{xy}>0$ and $S_x >0$ but $\Delta_{yz}$ is very small (Fig. 3 (e), (f)). It is clear applying this confinement and increasing the chemical potential leads to a phase transition from the uniaxial phase to the biaxial one. We found qualitative good agreement between our results and the achieved results from Onsager theory.[30] It is noteworthy we found similar results for uniaxial ellipsoidal particles with the aspect ratios *2a/2b*= 2.2 and 5.

Ordering properties of two-dimensional (2D) and quasi-two-dimensional (q2D) non-spherical colloids has been studied experimentally and theoretically.[31-36] The studies show there are differences between the nature of 2D nematic ordering and 3D one. So, we examined the orientation ordering properties of 2D hard ellipsoids. In this case, we confined these particles between two parallel hard walls which wall separations are considered smaller than the major axis of ellipsoids. As shown in Fig. 4, in 2D the U-B transition will form by increasing the chemical potential, and the number of particles that tend to be aligned with the *x* direction increases similar to the 3D slit pore. In this case increasing the chemical potential leads to the order parameters get very close to each other, i.e., $\Delta_{xy}=S_x$. Our results are in good agreement with the results of studying monolayers of hard ellipsoids by Varga et al.[36] They obtained a phase transition from a planar uniaxial nematic order to a biaxial nematic one.

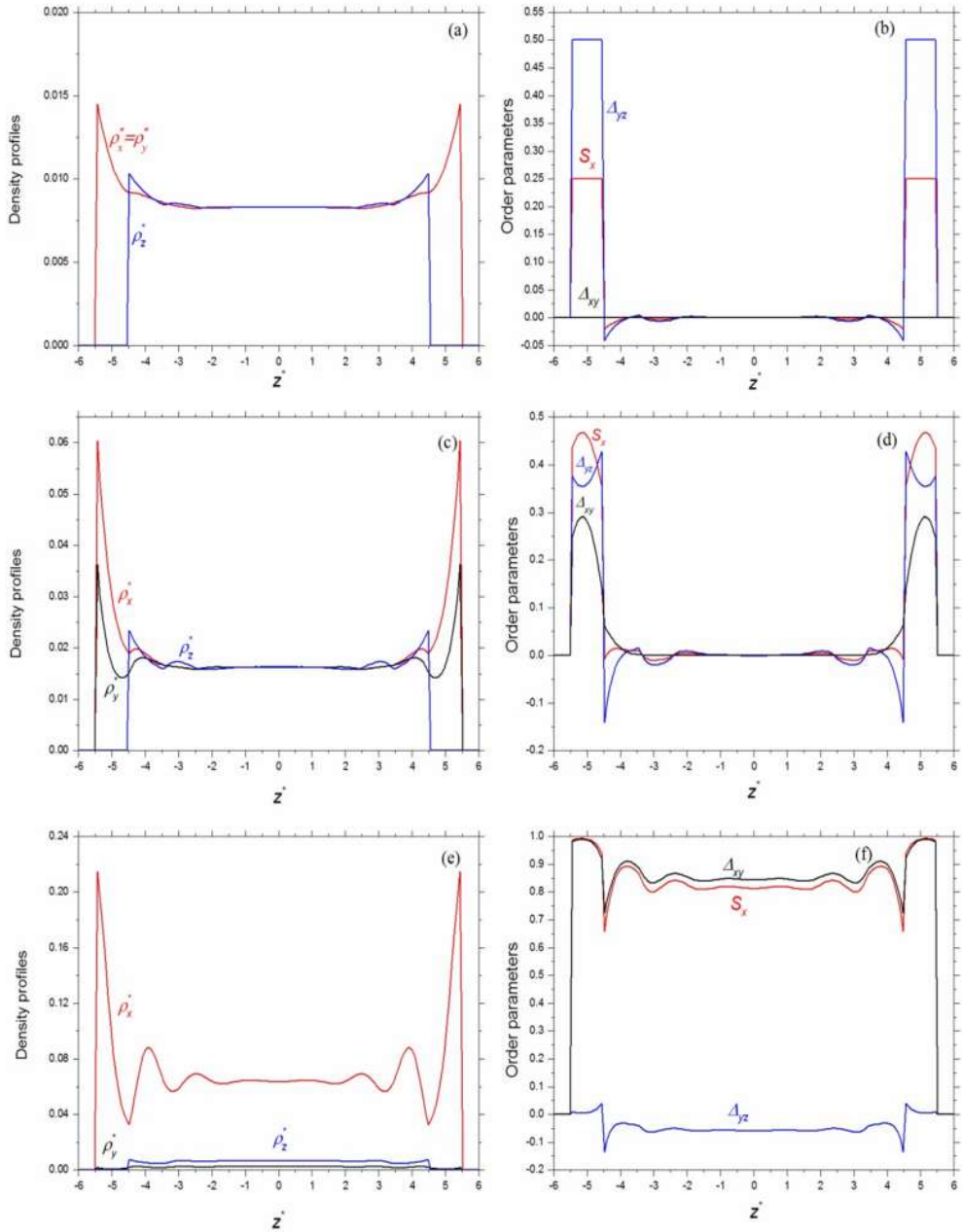

Fig. 3. Density profiles and order parameters of the hard ellipsoids (elongation $2a/2b=3$) confined between two hard walls at the wall separation of $h^*$ ($=h/2b$) $=12$. (a) and (b) obtained for the reduced chemical potential $\beta\mu=-3$, where the phase is called isotropic as in the middle of the pore all different axes have the same densities. (c) and (d) at $\beta\mu=-2.33$, this phase is biaxial near the walls but isotropic in the middle of the pore. Therefore the phase is called biaxial isotropic. (e) and (f) are plotted at $\beta\mu=-2.1$, where the phase is biaxial nematic and most of the particles are aligned with the x-axis even in the middle of the pore that is due to increasing the chemical potential or equivalently, the density.

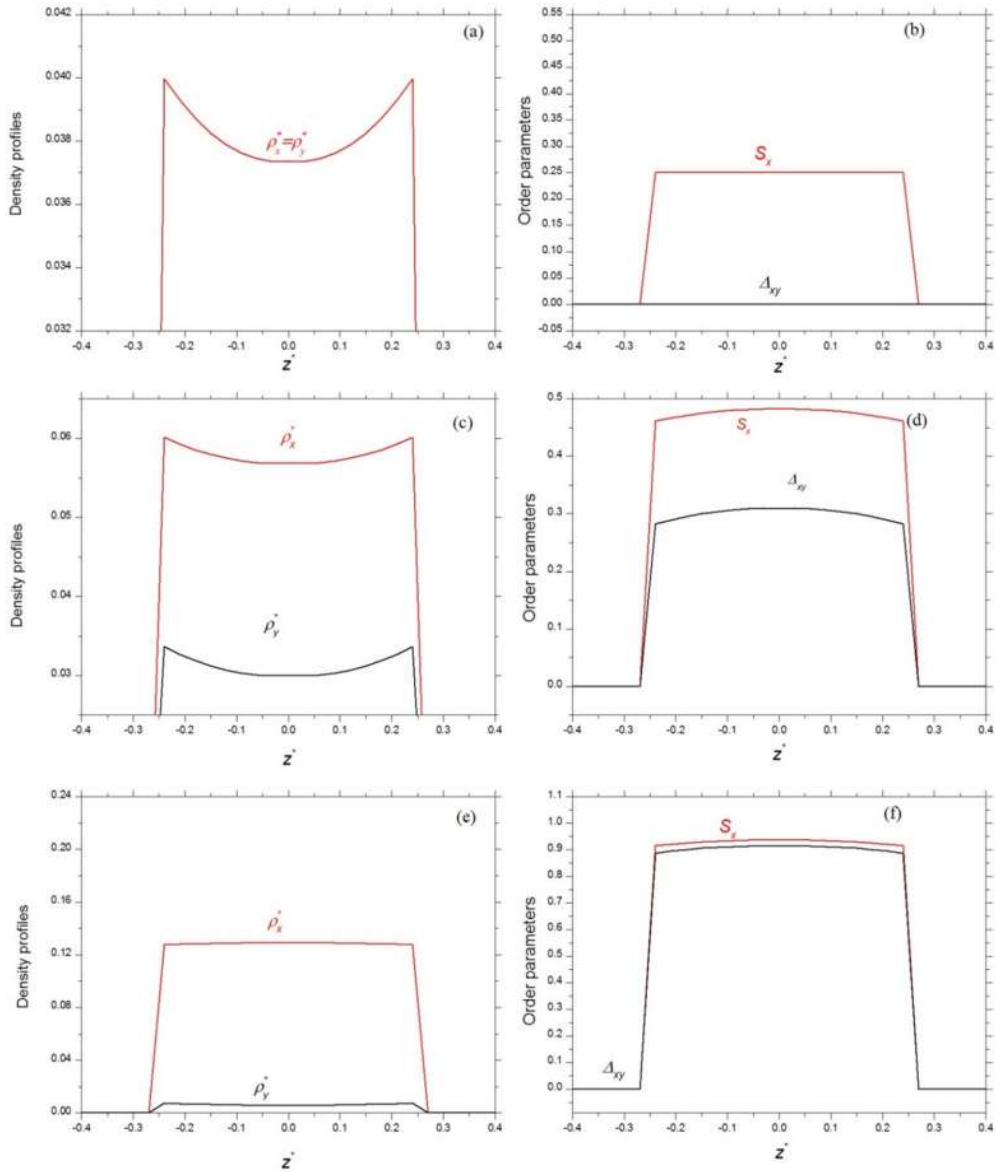

Fig. 4. Density profiles and order parameters of the quasi-2D system, where the wall separation is smaller than the aspect ratio of the particles ($2a/2b=3$ and $h^*=h/2b=1.5$). (a) and (b) are plotted for the reduced chemical potential $\beta\mu=-3$, where the phase is isotropic and there is no difference between the x and y-axes. (c) and (d) at $\beta\mu=-2.415$, which show there is also U-B phase transition even in the 2D systems like 3D one. (e) and (f) are obtained at $\beta\mu=-2.3$, where the phase is biaxial nematic.

## 4. Conclusion

The HNC density functional theory is used to consider hard ellipsoids confined between two parallel hard walls. At the first step, the excess grand potential with respect to its bulk as a unique functional of density was introduced. The DCF of homogeneous hard ellipsoidal fluid is the main required input. The minimization of the excess grand potential with respect to the density gave us some coupled integral equations to obtain the number density and the order parameters. We also used ROM model to find the density profiles and order parameters in contact with a single hard wall as reported in [27,28] for the hard rectangular rods. The presence of single wall leads to form a nematic layer near the wall that its thickness increases with increasing the chemical potential. It means there is a strong adsorption between the wall and the molecules. Then we obtained the density profiles and corresponding order parameters between two parallel hard walls. Because of the confinement, a phase transition from the uniaxial phase to the biaxial one at the larger chemical potential happens where the nematic film exists even throughout the pore. Similar behaviors have been also seen in the study of the hard rectangular rods [27, 28, 29]. We plan to investigate the phase transition diagram of the studied system which we expect to see a first order transition for the isotropic-nematic phase transition including a biaxial nematic phase and a continuous phase transition for the U-B one similar to [27,28]. It is good to mention here we have studied the equilibrium physical properties, however, some nonequilibrium behaviors can be examined as it has been done for liquid-vapour separation in [42].

**Appendix: Direct correlation function of hard ellipsoids**

The DCF of hard ellipsoids which are required in the mentioned equations has been calculated by Allen et al.[37] They used Monte Carlo simulation and found the DCF of hard ellipsoids and compared with Marko's results,[38] the results are in agreement. Here we use the improved Pynn-Wulf [39,40] expression for the DCF of hard ellipsoids proposed by Marko:

$$C(\vec{r}_{12}, \omega_1, \omega_2) = C_{PY}\left(\frac{|\vec{r}_1 - \vec{r}_2|}{\sigma(\hat{n}, \omega_1, \omega_2)}\right)[1 + \alpha P_2(\hat{\omega}_1.\hat{\omega}_2)] \quad (1A)$$

where $P_2(\mu) = (3\mu^2 - 1)/2$ and $\alpha$ are obtained by the same procedure as proposed by Marko, $C_{PY}$ is the Percus-Yevick DCF for the hard spheres,[1] $\sigma(\hat{n}, \omega_1, \omega_2)$ is the closest approach of hard ellipsoids. The modified closest approach introduced by Rickayzen [41] is given by:

$$\sigma(\hat{n},\omega_1,\omega_2) = 2b[1 - \chi \frac{(\hat{r}_{12}.\hat{\omega}_1)^2 + (\hat{r}_{12}.\hat{\omega}_2)^2 - 2\chi(\hat{r}_{12}.\hat{\omega}_1)(\hat{r}_{12}.\hat{\omega}_2)(\hat{\omega}_1.\hat{\omega}_2)}{1 - \chi^2(\hat{\omega}_1.\hat{\omega}_2)^2}$$
$$+ \frac{\lambda\left[(\hat{r}_{12}.\hat{\omega}_1)^2 - (\hat{r}_{12}.\hat{\omega}_2)^2\right]^2}{1 - \chi^2(\hat{\omega}_1.\hat{\omega}_2)^2}]^{-1/2} \quad (2A)$$

where:

$$\chi = \frac{a^2 - b^2}{a^2 + b^2} \quad , \quad \lambda = 1 - \chi\frac{4b^2}{(a+b)^2} \quad (3A)$$

parameters *2a* and *2b* denote the lengths of the major and minor axes of the ellipsoids, respectively.